\documentclass[prb,twocolumn,showpacs]{revtex4}
\usepackage{amssymb}
\usepackage {longtable}
\usepackage{amsmath}
\usepackage{graphics}
\usepackage{graphicx}
\usepackage{dcolumn}

\begin{document}

\title{THEORETICAL CONSIDERATIONS ABOUT THE GENERATION AND PROPERTIES OF
NARROW ELECTRON FLOWS IN SOLID STATE STRUCTURES}

\author{R.N.Gurzhi}
\author{A.N.Kalinenko}
\author{A.I.Kopeliovich}
\author{A.V.Yanovsky}
\affiliation{B.Verkin Institute for Low Temperature Physics \&
Engineering,  National Academy of Sciences of Ukraine, 47 Lenin
Ave, Kharkov, 61103, Ukraine}

\author{E.N.Bogachek}
\author{Uzi Landman}
\affiliation{School of Physics, Georgia Institute of Technology,
Atlanta, GA 30332-0430, USA}

\begin{abstract}
A method for the evaluation of the angular width of an electron
beam generated by a nanoconstriction is proposed and demonstrated.
The approach is based on analysis of a narrow-width electron flow,
that quantizes into modes inside a confining constriction which is
described in the adiabatic approximation, evolving into a freely
propagating electronic state after exiting the constriction. The
method that we developed allows us to find the parameters and the
shape of the constriction that are optimal for generation of
extremely narrow electron beams. In the case of a constriction
characterized by a linear widening shape an asymptotically exact
solution for the injection problem is found. That solution
verifies semi-quantitative results related to the angular
characteristics of the beam, and it opens the way for
determination of the distribution function of the electrons in the
beam. We have found the relationship between the angular
distribution of the electron density in the beam and the quantum
states of the electrons inside the constriction. Such narrow
electron beams may be employed in investigations of electronic
systems and in data manipulations in electronic and spintronic
devices.
\end{abstract}

 \maketitle

\section{INTRODUCTION}

Microconstrictions (referred to also as point contacts) connecting
macroscopic reservoirs are of particular interest in efforts aimed
at generation and investigation of ballistic quasiparticle
transport in solids \cite{al}. Recently, the development of
methods for imaging electron flows attracted significant attention
\cite{a2,a3,a4,a5,a6,a7,a8} due to it's potential to unveil the
details of electron motion in low-dimensional systems and to
provide insights into the behavior of devices in the quantum
regime. Moreover, with the use of a most recently developed
erasable electrostatic lithographic technique \cite{a9}, creation
of quantum constrictions with desired shapes has been
demonstrated. Additionally, metallic nanowires with high carrier
density \cite{b0} may also hold some promise as devices for
injection of electron flows. In light of above, the problem of
determining the operational parameters of an electron beam
injected through a constriction with a highly reduced size, is
both timely and important.

An electron flow injected through a constriction is in general
anisotropic. One of the first demonstrations of the importance of
the velocity anisotropy in electron flows can be found in
experiments with electron beams injected by quantum point contacts
\cite{b1,b2}, where a collimation effect \cite{b3} was found (see
also Ref.\onlinecite{b4}). The relative angular narrowness of an
electron beam allows experimental determination of the
electron-electron relaxation time \cite{b5,b6,b7,b8}. In the
scattering spectroscopy method proposed and demonstrated in
Ref.\onlinecite{b9} the narrowness of the electron beam plays a
key role: that is, the ability to control the scattering angle by
means of a narrow-angle beam injector, as well as a detector,
allows one to determine experimentally the electronic
angle-dependent differential scattering cross-sections associated
with different types of scatterers. Consequently, a narrow
electron beam may serve as a powerful tool for studying the
properties of electron scattering processes, and for determination
of the characteristics of the electron gas.

Narrow electron beams may also serve as a most effective tool for
the transmission of information in micro- and nano-devices
(including transportation of spin-polarized states \cite{c0}), and
as an instrument for handling the spin and the charge states of
quantum memory cells. In this context we remark that issues
pertaining to the angular and spatial distribution of narrow
electron beams are of great significance for the development of
high-resolution experimental techniques that utilize such beams,
as well as for the development and application of accurate
spatially targeted transfer of information using narrow electron
flows. We note here that, to date, the smallest angular width of
an electron beam injected into a two-dimensional electron gas
(2DEG) by a quantum point contact is of the order of $10^{o}$; in
Ref. \onlinecite{b9} an angular width $\Phi \approx 12^{o}$ was
observed (while Refs.\onlinecite{a2} and \onlinecite{a3} reported
a width $\Phi \approx 6^{o}$, it corresponds only to the most
pronounced central part of the electron flow).

\begin{figure}
\includegraphics[height=6 cm,width=6 cm]{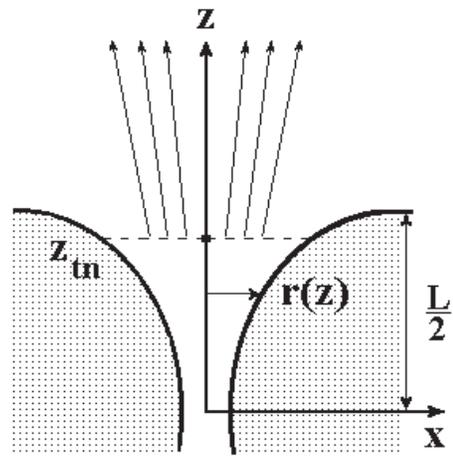}
\caption{Schematic of the constriction and an injected beam. The
length of the constriction $L$ is taken such that the detachment
point $z_{tn}$ is located inside the constriction.} \label{fig1}
\end{figure}

The main goal of our work is to analyze issues pertaining to the
prospect of generating super- narrow electron beams. To this end
we study also the distribution function of electrons in the beam,
since it enters considerations related to the selection of
conditions for formation of narrow beams. The interest in
conductance quantization in quantum two- and three-dimensional
constrictions (such as point contacts, nanowires and atomic
chains) \cite{c1,c2,c3}, \cite{c5} led to intensive investigations
of the electronic states in these systems. One of the main
characteristics of this phenomenon relates to the fact that the
quantized staircase-like variation of the conductance (with gate
voltage or constriction width) is determined by the adiabatic
properties of the constriction, and it is rather insensitive to
details of the geometrical configuration; here, ``adiabatic''
means a slow dependence of the constriction width $2r$ on the
coordinate $z$ along the longitudinal axis of the constriction
(see Fig.\ref{fig1}). The width changes noticeably on a scale that
exceeds essentially the minimal width $r(0)$ (see,
Ref.\onlinecite{c1}). However, the problem of the states of
electrons that have passed thought the constriction has not been
solved in the general case of the adiabatic approximation, since
the transformation of the adiabatic quantum states inside the
constriction to the distribution of freely moving electrons occurs
in a region where the adiabatic approximation ceased to be valid.
Nevertheless, in Ref.\onlinecite{b3} the characteristics of an
electron beam injected by a constriction have been studied in the
adiabatic approximation using the classical adiabatic invariant $I
= p_{x} \left( {z} \right)r\left( {z} \right)$. Due to the
conservation of the adiabatic invariant $I$, the beam converges
(the flaring effect, \cite{b3}) with increasing $z$, and near the
exit of the constriction we have
\begin{equation}
\label{eq1}
sin\left( {\frac{{\Phi} }{{2}}} \right) = \frac{{r\left( {0}
\right)}}{{r_{max}} },
\end{equation}
where $r_{max} $ is the half-width of the constriction at the
exit, and $r(0)$ is the half-width at $z=0$ (the origin of the $z$
axis is taken at the middle of the constriction). This result
\cite{b3} is valid, as will be shown in Section 1, only for
relatively ``short'' constrictions where the adiabatic
approximation is effectively valid for the entire constriction. A
simulation of the classical trajectories of the particles in such
constrictions has been presented in Ref. \onlinecite{b2}, and used
to determine the angular width of the beam.

In Section 1 we propose an approach that allows us to describe
qualitatively the motion of electrons exiting from the adiabatic
region and, thus, it permits analysis of the angular
characteristics of a beam injected by a constriction of an
arbitrary shape. In this case the parameters of the constriction
become particularly important at distances exceeding the
characteristic length-scale that determines the conductance
quantization behavior.

In Section 2 we find an asymptotically exact solution for electron
states in a constriction modelled by a linear widening. This
solution describes the conversion of adiabatic states inside the
constriction into states described by semi-classical wave
functions outside it, and it supports the results of the
qualitative study. The ``linear'' constriction that we study here
is also of additional interest since we find that in such a
constriction the pattern of the distribution of the electronic
density inside the constriction is maintained when the electrons
move away from the exit. Such distributions were observed in
Refs.\onlinecite{a4} and \onlinecite{a8} using scanning probe
microscopy (see also Ref.\onlinecite{a5} and references therein).

In Section 3 we consider the electronic distribution function of
the injected beam and compare our results with those of Refs.
\onlinecite{a4,a5,a6,a7,a8}, \onlinecite{b3} and \onlinecite{c7}.
We analyze the conditions when the distribution of electrons in
the beam reproduces the probability density function inside the
constriction; a distribution of this type has been observed in
Refs. \onlinecite{a4} and \onlinecite{a6,a7,a8}. We find also the
electron distribution in the opposite limiting case where the
constriction shape varies in a less smooth manner.

For the sake of simplicity we limit ourselves here to
two-dimensional constrictions, noting that the extension of our
results to the three-dimensional case is rather straightforward.
Additionally, we neglect electron-impurity scattering and consider
only the ballistic regime (which is readily achievable in 2D
heterostructure systems, see, e.g., Ref. \onlinecite{b2}). Because
of the scattering of electrons by the donor atom density
fluctuations (in 2D heterostructures) and by impurities \cite{a5},
the electron flow may form narrow branches with apparently small
changes in the total angular width of the flow. An additional
widening (spreading) of the electron flow $\Delta \Phi$ may be
estimated (in a diffusive approach) as $\Delta \Phi \sim \Phi_{0}
\sqrt{z/z_{0}}$, $z >> z_{0} $ (here $z$ is the distance along the
propagation axis from the point contact, $z_{0}$ is the mean scale
of the spatial fluctuations of the scattering potential, and
$\Phi_{0}$ is an average angular deviation of the electrons due to
the interactions with the fluctuations of the underliyng
potential). We remark that the distance dependence of the angular
widening of the beam caused by electron-electron interaction (see,
Ref.\onlinecite{c8}) is quite different from the above expression.

\section{INJECTION CONDITIONS FOR NARROW BEAMS}

Let us consider a constriction with an adiabatic narrow region;
apparently, other types of constrictions have been commonly found
to be unsuitable as effective injectors of narrow beams. Note that
the approach of Ref.\onlinecite{b3} which is based on employment
of an adiabatic invariant may be generalized to take into account
energy quantization in the constriction. It is known (see, for
example, Ref.\onlinecite{c9}) that in the semi-classical
approximation the adiabatic invariant is quantized in units of
$\hbar$. Qualitatively we may write for all the electron states in
the constriction
\begin{equation}
\label{eq2} I = p_{xn} \left( {z} \right)r_{n} \left( {z} \right)
\approx \hbar \left( {n + \gamma}  \right)\beta,
\end{equation}
where \textit{n}=1,2, ... is a discrete quantum number, $p_{xn}
\left( {z} \right)$ and $r_{n} \left( {z} \right)$ are the
root-mean-square values of $p_{x} $ and $x$, respectively, in the
n-th quantum state, and $\gamma$ and $\beta$ are numerical values
(of the order of unity) which depend on the model of the
confinement potential.

Let us show that the role of the breakdown of the adiabatic
approximation in the formation of a beam may be analyzed via the
use of a simple picture of ``detachment'' of the beam from the
constriction walls (at least for constrictions where the sign of
the wall curvature remains the same throughout). Detachment of the
beam occurs when the opening angle of the particles in the
constrictions (of the order of $p_{xn} \left( {z} \right)/p_{zn}
\left( {z} \right)$, that decreases with the distance from the
center due to the increase of $r_{n} \left( {z} \right)$) becomes
smaller than the corner angle of the constriction $dr_{n} \left(
{z} \right)/dz$. Thus, the ``detachment point'' $z_{tn} $ (see
Fig.{\ref{fig1}}) for the n-th mode of the beam may be determined
from the following equations
\begin{equation}
\label{eq3}
p_{zn} \left( {z} \right)r_{n} \left( {z} \right)\frac{{dr_{n} \left( {z}
\right)}}{{dz}} = \hbar \left( {n + \gamma}  \right)\beta ,
\end{equation}

\begin{equation}
\label{eq4}
p_{zn} \left( {z} \right) = \sqrt[]{2m\left( {\varepsilon _{F} - \varepsilon
_{n} \left( {z} \right)} \right)}
\end{equation}
Here, $\varepsilon _{n} \left( {z} \right)$ and $p_{zn} \left( {z}
\right)$ are, respectively, the energy of transverse motion and
the \textit{z} component of the momentum, which are well-defined
values in the adiabatic approximation \cite{c1}, $m$ is the
effective mass, and $\varepsilon _{F} $ is the Fermi-energy of the
electrons in the wide region; we assume that the voltage drop
across the constriction is small enough, that is $eV < <
\varepsilon _{F} $. The condition of the reality of $p_{zn} \left(
{0} \right)$ determines the number $n_{max} $ associated with the
last mode which can pass through the constriction. The angular
size $\Phi _{n} $ of the $n$-component of the beam is given by
\begin{equation}
\label{eq5}
sin\left( {\frac{{\Phi _{n}} }{{2}}} \right) \approx \frac{{\hbar \left( {n
+ \gamma}  \right)\beta} }{{p_{F} r_{n} \left( {z_{tn}}  \right)}},
\end{equation}
where $p_{F} = \sqrt{2m\varepsilon _{F}} $. This equation takes
into account possible variation of $p_z$ due to variation of the
confinement potential $U(x,z)$ at $z
> z_{tn} $.

Let us show next that the ``detachment point'' $z_{tn} $,
determined by Eqs. (\ref{eq3}) and (\ref{eq4}), coincides with the
limit of validity of the adiabatic approximation. The wave
function of an electron in the adiabatic approximation has the
following form $\psi = \eta _{n} \left( {x;z} \right)\varphi _{n}
\left( {z} \right)$ (see, Ref.\onlinecite{c1}), where the function
$\eta _{n} \left( {x;z} \right)$ satisfies the Sch\"{o}dinger
equation that is local with respect to $z$
\begin{equation}
\label{eq6}
\left( { - \frac{{\hbar ^{2}}}{{2m}}\frac{{\partial ^{2}}}{{\partial x^{2}}}
+ U\left( {x,z} \right)} \right)\eta _{n} = \varepsilon _{n} \left( {z}
\right)\eta _{n} .
\end{equation}
The function $\varphi _{n} \left( {z} \right)$ is the wave
function associated with longitudinal motion (along the axis of
the constriction) in the field of the ``effective potential''
$\varepsilon _{n} \left( {z} \right)$. From examination of the
terms in the complete Schr\"{o}dinger equation that are maintained
in comparison with those that are omitted in the adiabatic
approximation (these include the terms $\varphi \partial ^{2}\eta
/\partial z^{2}$ and $\left( {\partial \eta /\partial z}
\right)\left( {\partial \varphi /\partial z} \right)$), we obtain
the following inequalities (in Eq.(\ref{eq7}) primes denote
derivatives with respect to z)
\begin{equation}
\label{eq7}
nr_{n}^{'2} ,r_{n} r_{n}^{''} ,\frac{{r_{n} p_{zn} r_{n}^{'}} }{{\hbar} } <
< n.
\end{equation}
These inequalities determine the region where the adiabatic
approximation is valid. It is easy to check that the last
inequality will break down first (or simultaneously with the
others) when $z$ increases ($z > 0$). To prove this, it is enough
to consider the region where $r_{n} \left( {z} \right) - r_{n}
\left( {0} \right) > r_{n} \left( {0} \right)$, because in this
narrow region the validity of all these inequalities is equivalent
to the initial assumption about the adiabatic constriction. If we
assume that $r_{n} $ increases monotonically with the increase of
the z-coordinate and that $U(x,z)$ decreases monotonically (and,
therefore, $\varepsilon _{n} \approx p_{xn}^{2} /2m + U\left(
{0,z} \right)$ decreases too), it follows from Eq.(\ref{eq4}),
that $p_{zn} \ge p_{xn} \approx \hbar n/r_{n} $ for modes which
move through the constriction, thus proving our conjecture.
Therefore, the regions that are associated with the adiabatic
approximation and with free propagation of the particles are
adjacent to each other, and there is no intermediate asymptotic
region between them. This conclusion justifies our suggestion that
the opening angle of the constriction $\Phi = \Phi _{n_{max} } $
could be evaluated from Eqs.(\ref{eq3} - \ref{eq5}).

To end our discussion of Eq.(\ref{eq7}) we note that the validity
of the inequalities $r'' > > n/r > > r'p_{z} /\hbar $ may be
extended to the case that the profile of the constriction has a
``break'', i.e. a small region with a large shape-curvature. If
$r'<<1$ on both sides of the break it leads to only small
corrections to the electron wave functions. Imperfections in the
profile of the constriction (such as breaks or steps) which are
small compared with the electron wave length have only a weak
effect on the characteristics of the beam.

In the hard - wall model that we mainly use below, $r_n(z)$ does
not depend on $n$ and it is equal to the half-width of the
constriction $r(z)$. Also, $\varepsilon _{n} \left( {z} \right) =
\left( {\pi \hbar n/2r\left( {z} \right)} \right)^{2}/2m + U\left(
{z} \right)$, where $U(z)$ is the part of the potential that
depends on the $z$-coordinate, $\gamma=0$, $\beta=\pi/2$. We
analyze first the possibility of generating a narrow beam in a
constriction with no potential barrier in the center, i.e.
$U(z)=0$. In this case, $n_{max} \approx 2p_{F} r\left( {0}
\right)/\pi \hbar $ and we obtain from Eq.(\ref{eq5})
\begin{equation}
\label{eq8}
sin\left( {\frac{{\Phi} }{{2}}} \right) = \frac{{r\left( {0}
\right)}}{{r\left( {z_{tn_{max}} }  \right)}}.
\end{equation}
Note that Eq.(\ref{eq8}) is similar to Eq.(\ref{eq1}) of
Ref.\onlinecite{b3}, with the only distinction regarding the
occurance of $r\left( {z_{tn_{max}} }  \right)$, instead of
$r_{max} $. Since we consider here a narrow beam, $\Phi << 1$, in
order to find the detachment point we may analyze Eq. (\ref{eq3})
far away from the center of the constriction, where $r(z) >> r(0)$
and where, following Eq.(\ref{eq4}), $p_{zn} \approx p_{F} $. Let
the shape of the constriction in this region be given by the
following power dependence: $r\left( {z} \right) = a|z|^{\alpha}
$; from the evident condition $r\left( {z_{tn_{max}} }  \right) >
> r\left( {0} \right)$ we readily conclude that $a < < r\left( {0}
\right)^{1 - \alpha} $. Consequently, from Eq.(\ref{eq3}) and the
aforementioned estimate for $n_{max} $, we obtain that in order to
achieve the minimal angular width Ö the constriction length L (see
Fig.\ref{fig1}) should be made approximately equal to
$z_{tn_{max}}$
\begin{equation}
\label{eq9}
 L \approx z_{tn_{max}}, \ \mbox{where}\ z_{tn_{max}}  \approx 4\alpha \frac{{r\left(
{0} \right)}}{{\Phi ^{2}}}.
\end{equation}
If the length of the constriction, $L$, is less than $z_{tn_{max}}
$, the resulting angular width $\Phi$ increases and is given by
Eq. (\ref{eq1}), while for $L > z_{tn_{max}}$ the angular width of
the beam is unaffected and it remains as given in Eq.(\ref{eq9}).
In other words, to generate a flow with an angular width $\Phi$
one may need to use a constriction with an effective length that
is not smaller than $z_{tn_{max}}$, as determined in
Eq.(\ref{eq9}). Therefore, we conclude that the ``flaring effect''
\cite{b3} produces narrow beams only for relatively long
constrictions.

Decreasing the relative length of the constriction is related to a
decrease of the exponent $\alpha$. It is evident that the
detachment of a beam is possible only if $\alpha>1$. Nevertheless,
if $1/2 < \alpha < 1$, the condition $z < < z_{tn_{max}} \approx
\left( {r\left( {0} \right)/a^{2}} \right)^{1/\left( {2a - 1}
\right)}$ determines the adiabatic region. At $z > > z_{tn_{max}}
$ the propagation of the electrons can be described in terms of
classical mechanics. It is possible to verify that Eq.(\ref{eq9})
remains valid in this case and that the optimal length of the
constriction (required in order to generate a narrow beam) can be
estimated to be of the order of $z_{tn_{max}} $.

\begin{figure}
\includegraphics[height=3 cm,width=6 cm]{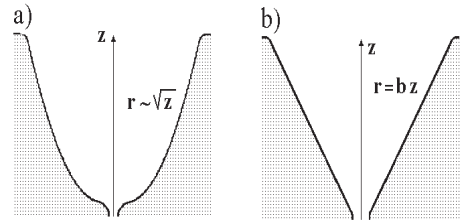}
\caption{Constrictions of different shapes: (a) a parabolic
constriction, with $r^{2} \approx r\left( {0} \right)z$ at
$r>>r(0)$, and (b) a linear widening constriction. } \label{fig2}
\end{figure}

The case when $\alpha=1/2$ is of special interest. When $a^{2} =
2r$ and $p_{z} \approx p_{F} $ Eq.(\ref{eq3}) can be used for all
values of $z$, and the adiabatic condition is fulfilled everywhere
in the constriction. Thus, for $\alpha=1/2$ Eq.(\ref{eq9}) is
valid for any length of constriction (if $\Phi<<1$). This differs
from the case of $\alpha > 1/2$, where, as aforementioned, an
increase of $L$ beyond the detachment point $z_t$ does not reduce
the angular width of the beam. When $\Phi << 1$, see
Eq.(\ref{eq9}), $L \approx 2r\left( {0} \right)/\Phi ^{2}$ (at
$a^{2} \approx r\left( {0} \right)$) will be valid for arbitrary
length of the constriction. In the case where $\alpha < 1/2$ the
relation between the relative length and the angle $\Phi$ is less
favorable in the adiabatic region $z
> > z_{tn_{max}} $. Therefore, a constriction of parabolic shape, $r^{2}
\approx r\left( {0} \right)z$ (see Fig.\ref{fig2}), is the optimal
choice. The case when $\alpha = 1$ will be discussed in details in
the next section.

For a model of a ``square'' constriction \cite{c1} $r = r\left(
{0} \right) + 2z^{2}/R$, with $r(0)<< R$, and from Eqs.(\ref{eq8})
and (\ref{eq9}) we obtain for $\Phi << 1$
\begin{equation}
\label{eq10} \Phi \approx 4\left( {r\left( {0} \right)/R}
\right)^{1/3}, \quad L \approx \left( {1/2} \right)\left( {r\left(
{0} \right)R^{2}} \right)^{1/3}.
\end{equation}
From this expression we conclude that the distance scale for
formation of an electron beam is larger than the distance (of the
order of $\left( {r\left( {0} \right)R} \right)^{1/2}$) that
determines the conductance quantization.

The potential barrier in the center of constriction may also lead
to narrowing of the electron flow \cite{b3}. The cause is that in
addition to the flaring effect with increasing $z$, the $p_z$
component of the momentum increases also due to the influence of
the potential $U(z)$.

In the hard wall approximation we may write Eqs.(\ref{eq3}) - (\ref{eq5}) for $n = n_{max} $
in the following form
\[
p_{F} \left( {z_{t}}  \right)r\left( {z_{t}}  \right)\frac{{dr}}{{dz}} =
p_{F} \left( {0} \right)r\left( {0} \right),
\]

\begin{equation}
\label{eq11} p_{F} \left( {z} \right) = \sqrt[]{2m\left(
{\varepsilon _{F} - U\left( {z} \right)} \right)}
\end{equation}

\[
\Phi \approx 2\frac{{p_{F} \left( {0} \right)r\left( {0} \right)}}{{p_{F}
r\left( {z_{t}}  \right)}}.
\]

Here we assume also that $\Phi<<1$ and $p_{z} \left( {z_{t}}
\right) \approx p_{F} \left( {z_{t}}  \right)$. As may be seen
from Eqs.(\ref{eq11}), the flaring effect and the effect of the
potential are independent from each other only when $U(z)=const$
at $z < z_t$; otherwise the potential barrier leads to a reduction
of $r\left( {z_{t}} \right)$, i.e. it results in an attenuation of
the flaring effect. Thus, in the case of a linear constriction,
i.e. $r \propto z$, the two effects will compensate each other (if
$U(z)=0$ at $z > z_t$); the opening angle does not vary when the
potential is switched on, but the optimal relative length, $L
\approx z_{t} $, is reduced.

An alternative way to obtain a narrow beam, without having to
resort to the use of a long constriction, consists of the
application of an added repulsive potential. For a sufficiently
wide constriction ($r\left( {0} \right) > > \lambda _{F} \equiv
2\pi \hbar /p_{F} $ and a length that exceeds slightly the width)
it is sufficient to apply a potential that is transparent for one
mode ($n = 1$) only, i.e. $\varepsilon _{F} - U\left( {0} \right)
= \left( {\pi \hbar /r\left( {0} \right)} \right)^{2}/8m$. From
Eq.(\ref{eq1}), we obtain an opening angle $\Phi \approx \lambda
_{F} /r\left( {0} \right)$ (for short constriction $r\left(
{z_{t}} \right) \approx r\left( {0} \right)$).

Note that Eqs.(\ref{eq8} - \ref{eq11}) do not include the Planck
constant - indeed, they use only a classical adiabatic invariant
and classical considerations pertaining to the breakdown of
adiabaticity (the detachment of the beam). But, if we would like
to minimize both the angular and spatial width (that is the
transverse size) of the beam near the exit from the constriction
we have to take into account the minimal product of these values,
$r\left( {z_{tn_{max}} }  \right)\Phi \approx \lambda _{F} $,
allowed by the uncertainty principle. This underlies the finding
that in order to obtain an ``integrally'' narrow beam one has to
use a metallic with a small electron wave-length at the Fermi
level. Here an ''integrally'' narrow beam means an electron flow
with both the transverse width of the flow and the angular
spreading restricted to small values.

\section{BEAM INJECTION BY A LINEAR SHAPE CONSTRICTION}

Let us consider here the electron states in a constriction
characterized by a linear-widening shape (see Fig.{\ref{fig2}b},
i.e. $r={bz}$ at $r >> r(0)$. We show below that when $b << 1$
this problem has a simple, and an asymptotically exact, solution.
Note that a constriction with a linear widening shape is a special
case of a hyperbolic constriction. In this case the variables in
the Schr\"{o}dinger equation can be separated, thus allowing one
to obtain a solution for the conductance in this type of contacts
\cite{c3}.

We use the aforementioned fact that $p_x$ decreases in an
adiabatic widening when the electron propagates from $r(0)$ to $r
>> r(0)$. This underlies the validity of the inequalities
$p_{x} < < p \equiv \sqrt{2m\varepsilon} $ and $(p-p_z) << p$. The
electron wave function may be written in the form
\begin{equation}
\label{eq12}
\Psi \left( {x,z} \right) = \psi \left( {x,z}
\right)exp\left( {i\frac{{pz}}{{\hbar} }} \right).
\end{equation}
Using the hard - wall model in the linear section of the
constriction and taking into account that the value of the
component $p_z$ is close to the whole momentum $p$, we may neglect
in the Schr\"{o}dinger equation the second derivative of $\psi$
with respect to $z$
\begin{equation}
\label{eq13} \left( {\frac{{\hbar ^{2}}}{{2m}}}
\right)\frac{{\partial ^{2}\psi }}{{\partial x^{2}}} +
i\frac{{\hbar p}}{{m}}\frac{{\partial \psi }}{{\partial z}} = 0.
\end{equation}
It is readily observed that the solutions of Eq. (\ref{eq13}) with
a vanishing boundary condition, $\psi \left( {\left| {x} \right| =
r\left( {z} \right),z} \right) = 0$, have the following form
\begin{equation}
\label{eq14} \psi_{n} = \left\{
\begin{array}{rl}
 \frac{1}{\sqrt{bz}} \sin{ \frac{ \pi n}{2} \bigl(
 \frac{x}{bz}+1 \bigr)}e^{\frac{ip}{2\hbar z} \bigl[ x^2+\bigl( \frac{\pi n
 \hbar }{2bp} \bigr) ^2 \bigr]},&  \ x < bz,\\
 0, &  \ x > bz.
\end{array} \right.
\end{equation}

Using these functions for estimations of the omitted term in the
Schr\"{o}dinger equation, we observe that our initial assumption
is valid if $b<<1$ and $z >> n\lambda _{F} /b$ (the omitted term
is less than the second one on the left-hand side of Eq.
(\ref{eq13})). Taking into account that $n_{max} \approx r\left(
{0} \right)/\lambda _{F} $ for electron modes passing thought the
constriction, we find that the last inequality is equivalent to
the condition $r >> r(0)$.

When $z < < n\lambda _{F} /b^{2}$, we can neglect the $x^2$
dependence of the exponent in Eq.(\ref{eq14}) compared with the
$x$ dependence of the trigonometric function and, consequently,
the wave function $\Psi_{n}$ has an adiabatic form \cite{c1}. If
$z>> n\lambda _{F} /b^{2}$ ($p \approx p_{F} = 2\pi \hbar /\lambda
_{F} $) the wave function in Eq.(\ref{eq14}) describes (in the
semi-classical approximation) a beam of quasi-particles (whose
distribution function we discuss in the next section) which
propagates freely inside a solid angle $\Phi=2arctan(b)$. In some
sense, the detachment of the beam from the side walls occurs also
in the linear constriction - here, when $z >> n\lambda _{F}
/b^{2}$ particles ``glide'' along the walls and thus one can
neglect their interaction with the walls. Therefore, the solution
given in Eq.(\ref{eq14}) allows us to trace the transformation of
the adiabatic modes inside the constriction to the beam states
described by the classical distribution function.

We remark that the limit of the adiabatic region found by us,
$n_{max} \lambda _{F} /4b^{2} \approx r\left( {0} \right)/\Phi
^{2}$, supports also the result given in Eq.(\ref{eq9}) of the
previous section. It is of importance that when $b<<1$, this limit
is placed in the domain of applicability of the solution given by
Eq.(\ref{eq14}), $r >> r(0)$. Thus, the solution in
Eq.(\ref{eq14}) can be matched with an adiabatic wave function
\cite{c1} that corresponds to small $z$, where the shape of the
constriction deviates from the linear form. Consequently, the
single inequality $b<<1$, permits us to describe analytically the
electron state for all values of the coordinate $z$.

Note also that a solution of the type given in Eq.(\ref{eq14}) may
be obtained in the ``soft'' - wall model for certain types of
potentials forming the constriction. Let us use in the following a
potential given by $U\left( {x,z} \right) = z^{ - 2}u\left( {x/z}
\right)$, and let $\eta _{n} $ denote the solutions of the
``local'' Schr\"{o}dinger equation with eigenvalues $\tilde
{\varepsilon} _{n} $
\begin{equation}
\label{eq15}
\frac{{\hbar ^{2}}}{{2m}}\eta _{n}^{''} + u\eta _{n}
= \tilde {\varepsilon }_{n} \eta _{n} .
\end{equation}
Here the derivatives are taken with respect to \textit{x/z}. An
equation similar to Eq.(\ref{eq13}) is given by
\begin{equation}
\label{eq16} \frac{{\hbar ^{2}}}{{2m}}\frac{{\partial ^{2}\psi}
}{{\partial x^{2}}} - U\left( {x,z} \right)\psi + \frac{{i\hbar
p}}{{m}}\frac{{\partial \psi }}{{\partial z}} = 0.
\end{equation}
The solutions of Eq.(\ref{eq16}) are
\begin{equation}
\label{eq17} \psi _{n} = \frac{1}{\sqrt{z}}\eta_n \Bigl(
\frac{x}{z} \Bigr) exp \Bigl( \frac{i}{\hbar z}\Bigl[
\frac{px^2}{2}+\frac{\tilde{\varepsilon}_n m}{p}\Bigr]\Bigr).
\end{equation}
For Eq.(\ref{eq16}) to serve as a good approximation to the
complete Schr\"{o}dinger equation, the conditions $z > >
\sqrt{\tilde {\varepsilon} _{n} m/p^{2}}$ and $b_n <<1$ have to be
fulfilled. An example where these conditions are fulfilled is
provided by the potential $U\left( {x,z} \right) = c\left(
{x^{2}/z^{4}} \right) + d/z^{2}$, where $c$ and $d$ are constants,
and $c >> (\hbar n)^{2}/m $. In the above, $b_n$ may be termed as
the ``localization radius'' of the functions $\eta_{n} $. For the
soft-wall potential discussed here, $b_n$ plays (for the nth-mode)
the same role as the parameter $b$ introduced earlier in the
context of the hard-wall model (see the beginning of this section,
Eq.(\ref{eq14})); physically, $b_n$ is the turning point in
Eq.(\ref{eq15}), corresponding to the location where $u(x/z)
=\tilde {\varepsilon}_{n} $ and consequently the kinetic energy
vanishes there -- we thus conclude that while for $x/z < b_n$ the
function $\eta_{n} $ takes finite values, it decreases (typically
exponentially) for $x/z > b_n$.

\section{THE DISTRIBUTION FUNCTION OF ELECTRONS IN A BEAM}

The wave length of the electron in the $x$-direction, $h/p_x$,
becomes less than the transverse size of the beam at a distance
(along the constriction axis) $z >> z_{tn_{max}} $ from the center
of the constriction. For such circumstances the electron beam may
be considered as a classical object, and the distribution function
of such a classical beam, radiated from a small region, may be
written as
\begin{equation}
\label{eq18}
f\left( {p_{x} ,x,z} \right) = \rho \left( {x,z}
\right)\delta \left( {p_{x} - xp/z} \right),
\end{equation}
\[
\rho \left( {x,z} \right) = z^{ - 1}\chi \left( {x/z} \right),
\]
where $\rho(x,z)$ is the distribution of the electrons with
coordinates $x$ and $z$; $p_{z} \approx p$ because the beam is
assumed to be narrow. We suppose also that all electrons in the
beam have a definite energy, $p^2/2m$. The function $\chi(\theta)$
is the angular distribution of particles, expressing the deviation
from the beam axis. The distribution in Eq.(\ref{eq18}) satisfies
the condition of conservation of the particle flow, \textit{i.e.}
$\int{ \rho \left( x,z \right) dx}=const$.

When $z > > z_{tn_{max}}  $ the exact solution given by
Eq.(\ref{eq14}) is the semi-classical wave function (the rapid
$x$-dependence is due to the $x^2$ term in the exponent,
$p_x=xp_z/z$) and it leads to the distribution function described
by Eq.(\ref{eq18}). The contribution of $n$-th mode to the
distribution function $\chi(\theta)$ (normalized to unity, i.e.
$\int {\chi _{n} \left( {\theta}  \right)d\theta = 1} $) has the
form
\begin{equation}
\label{eq19} \chi _{n} \left( {\theta}  \right) = z\left| {\Psi
_{n} \left( {x,z} \right)} \right|^{2}, \quad \theta = x/z,
\end{equation}

\begin{equation}
\label{eq20} \left| {\Psi _{n} \left( {x,z} \right)} \right|^{2} =
\left\{ {{\begin{array}{*{20}c}
 {{\begin{array}{*{20}c}
 {\left( \frac{1}{bz} \right)sin^{2}\left( {\frac{{\pi n}}{{2}}\left(
{\frac{{\theta} }{{b}} + 1} \right)} \right),} \hfill & {|\theta | < b,}
\hfill \\
\end{array}} } \hfill \\
 {} \hfill \\
 {{\begin{array}{*{20}c}
 {0}, \hfill & {|\theta | > b.} \hfill \\
\end{array}} } \hfill \\
\end{array}} } \right..
\end{equation}
Thus, in the linear constriction model, the density of particles
in the beam reproduces exactly the density of the corresponding
adiabatic mode. This is true also in the case of a constriction
modeled by ``soft'' walls (($\chi _{n} = |\eta _{n} |^{2}$, see
Eq.(\ref{eq17})).

The above demonstrates that the linear constriction model yields an
optimally ``smooth'' transition from the adiabatic states to the classical
ones when the pattern of the distribution of the electronic density inside
the constriction, $|\Psi _{n} \left( {x} \right)|^{2}$, is maintained as the
electrons move away from the exit

Let us consider now a constriction model that describes the
opposite limit to the linear constriction discussed above -- that
is, when the constriction ends abruptly in the adiabatic region
(this problem has been considered numerically in
Ref.\onlinecite{c7}). Note first, that Eq.(\ref{eq13}) is
equivalent to the one-dimensional time-dependent Schr\"{o}dinger
equation; the time of motion along the $z$-axis is $t = zm/p$.
Consequently, when $\Phi<<1$ the problem concerning the behavior
of particles leaving the adiabatic constriction can be mapped onto
the one concerning determination of the response of particles
initially localized in a potential well to the sudden removal of
the well. The latter problem has an evident solution -
\textit{i.e}., in the (momentum) $p_x$-representation, the density
$|\Psi _{n} \left( {p_{x}} \right)|^{2}$ (instead of $|\Psi _{n}
\left( {x/z} \right)|^{2}$, as was the case for the linear
constriction) is conserved in time. Taking into account
Eq.(\ref{eq18}) we obtain
\begin{equation}
\label{eq21} \chi _{n} \left( {q} \right) = 2\pi p\hbar \left|
{\Psi _{n} \left( {p_{x} = p\theta}  \right)} \right|^{2}.
\end{equation}
In the hard wall potential model
\begin{equation}
\label{eq22} \left| {\Psi _{n} \left( {p\theta}  \right)}
\right|^{2} = \frac{{n^{2}r_{t} sin^{2}\left( {kr_{t} + \frac{{\pi
n}}{{2}}} \right)}}{{4\left[ {\left( {kr_{t}}  \right)^{2} -
\left( {\frac{{\pi n}}{{2}}} \right)^{2}} \right]^{2}\hbar ^{2}}},
\quad k = \frac{{\theta p}}{{\hbar} },
\end{equation}
where $2r_t$ is the width of the constriction at the place where
the constriction terminates. The main difference between the
distributions given in Eqs.(\ref{eq19}), (\ref{eq20}) and Eqs.
(\ref{eq21}), (\ref{eq22}) is that in the first case the
distributions have the same angular size Ö for all $n$, while in
the second case the distributions are localized near the angles
$\theta = \pm \pi \hbar \left( {n - 1} \right)/2r_{t} p$ (the
width of the main peaks is of the order of $\hbar /r_{t} p$).

The function described in Eq.(\ref{eq22}) is valid for an
arbitrary shape of the constriction, if we interpret $\Psi _{n}
\left( {p_{x}}  \right)$ as the wave function of the electron at
the exit of the constriction ($z=z_t$, $p_x<<p_z$). While in
general this wave function differs from the one at the center of
the constriction, the two are similar when the electron does not
undergo any collisions with the walls after it leaves the
adiabatic region. The latter takes place when the radius of
curvature of the constriction in the ``detachment'' region
satisfies the condition $R < < r_{t} /\Phi ^{2}$-- this inequality
is the applicability condition of Eq.(\ref{eq21}). In the opposite
limiting case, i.e. for $R > > r_{t} /\Phi ^{2}$, Eq.(\ref{eq19})
is valid. Here the radius of the constriction at the detachment
point $r_t$ (where the adiabatic approximation is violated) can be
determined as the maximum value of $r$ in the region where $dr/dz
\approx \Phi $; $R$ is the radius of curvature of the constriction
in this region.


\begin{figure}
\includegraphics[height=6 cm,width=6 cm]{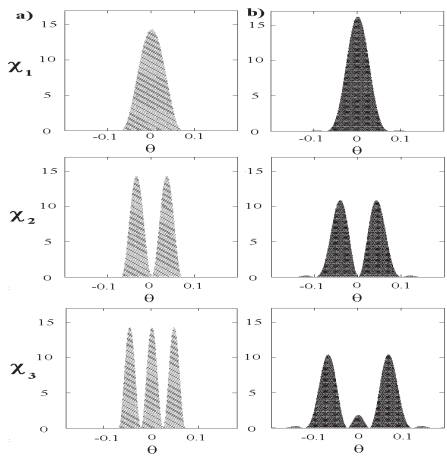}
\caption{The angular ($\theta$ in radians) distribution $\chi _{n}
$of the $n$-th mode\textit{} for n=1,2 and 3, plotted for: (a) a
constriction with a shape close to that with a linear widening,
and (b) a constriction that ends abruptly ($r_{t} /\lambda _{F} =
10$).} \label{fig3}
\end{figure}

\begin{figure}
\includegraphics[height=6 cm,width=6 cm]{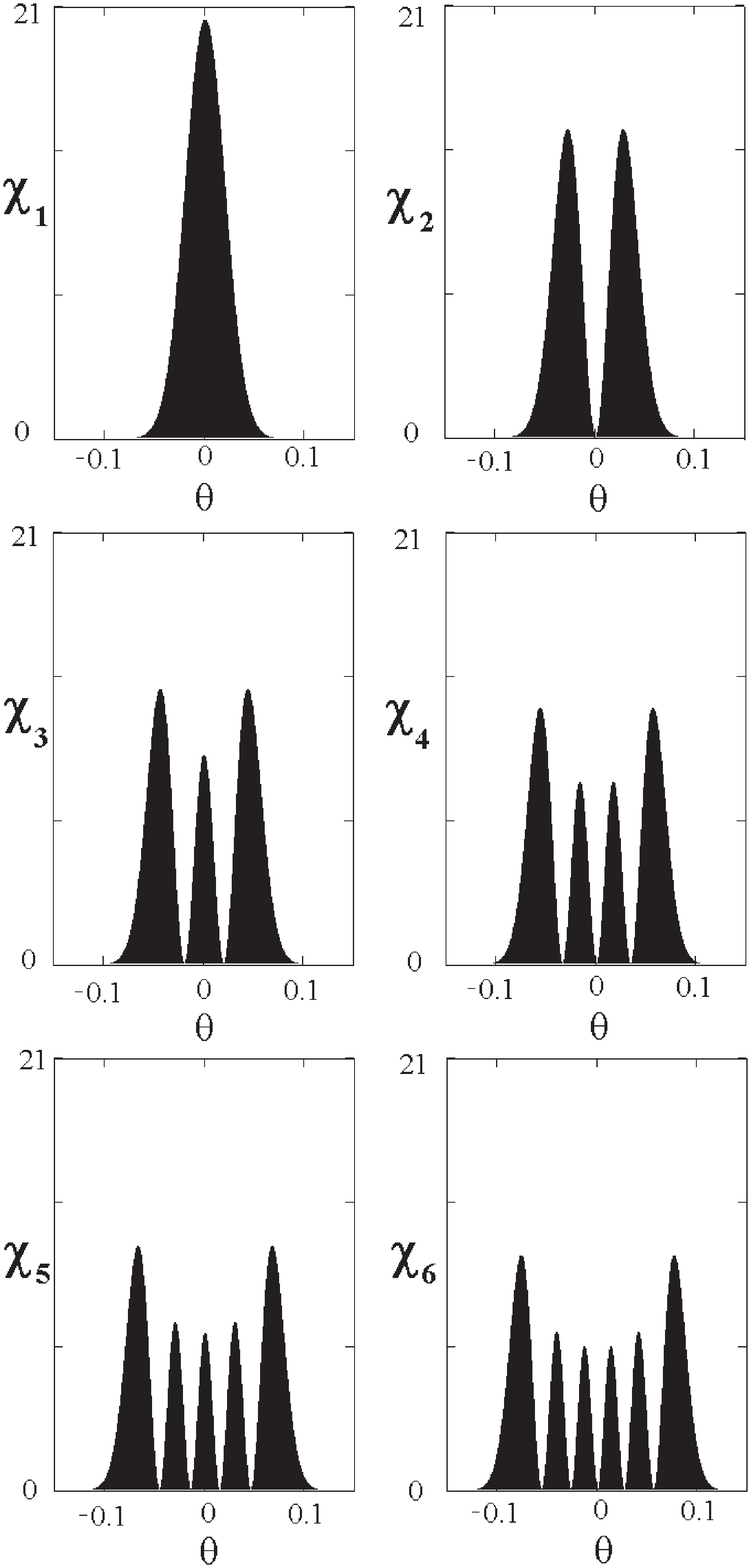}
\caption{The angular ($\theta$ in radians) distribution $\chi _{n}
$of the \textit{n}-th mode for \textit{n}=1,2,3,5 and 6, generated
by a soft-wall constriction.} \label{fig4}
\end{figure}

The $\theta$-dependencies of $\chi _{n} $ for the first three
quantum modes in the hard-wall constriction model are displayed in
Fig.{\ref{fig3}}. The electron modes radiated by a constriction
with a shape close to the linear widening one (a radius of
curvature $R >> r_{t} /\Phi ^{2}$) are displayed in Fig.3a. These
modes reproduce the $x$-dependence of the $|\psi |^{2}$ function
inside the constriction. In Fig.3b we display the radiation from a
constriction which ends abruptly in the adiabatic region, $R < <
r_{t} /\Phi ^{2}$. The difference between the characteristics of
the electron flows generated by the two types of constrictions is
evident (compare, in particular, the angular distributions for the
third mode). Note that in the model of a harmonic transverse
potential (soft-wall constriction model) the distributions are the
same for both types of constrictions. In this case, the wave
functions are the same in the coordinate and momentum
representations. We may define the angular width of the electron
beam by introducing the number of modes passing through the
constriction, $n_{max} $, and the maximal value of the
\textit{x}-component of the electron momentum at the detachment
point, $\Phi \approx p_{xt} /p_{F} \sqrt{2\left( {0.5 + n_{max}}
\right)}$. These values correspond to a definite value of the
coefficient in the transverse potential at the detachment point:
$c\left( {z_{t}}  \right) = p_{xt}^{4} /8m\hbar ^{2}\left( {0.5 +
n_{max}}  \right)^{2}$. The corresponding half-width of the
electron state in the detachment point is $r_{t} = 2\hbar \left(
{0.5 + n_{max}} \right)/p_{xt} $. The $\theta$-dependencies of
$c_n$ for the first six quantum modes in the soft-wall model are
displayed in Fig.{\ref{fig4}} where we have taken the same angular
width $p_{xt} /2p_{F} \sqrt[]{2\left( {0.5 + n_{max}}
\right)}\simeq 0.07$ as in Fig.{\ref{fig3}}. Similar
$\theta$-dependencies for the first three quantum modes in the
harmonic confinement potential model, have been discussed and
observed experimentally in Refs. \onlinecite{a3} and
\onlinecite{a4}. The half-width of the constriction at the
detachment point satisfies the equation $c\left( {z_{t}}
\right)r_{t}^{2} = p_{xt}^{2} /2m$ (this differs from the equation
$c\left( {0} \right)r^{2} = \varepsilon _{F} $ used widely for the
definition of the width of the constriction in the narrowest
region in soft-potential models).

Let us finally discuss the total electron flow injected by the constriction.
This flow is a sum over all the modes that pass through the constriction
\begin{equation}
\label{eq23} \chi = \frac{{VmG_{0}} }{{ep_{F}} }\sum\limits_{n =
1}^{n_{max}}  {} \chi _{n} ,
\end{equation}
where V is the potential difference between the two reservoirs
which are connected by the constriction, and $G_0=2e^2/h$ is the
conductance quantum. The coefficient in front of the summation is
chosen in order to maintain a well-known quantization rule for the
regime that is linear in V ,see Ref.\onlinecite{c6}. For a
sufficiently wide constriction, $r\left( {0} \right) > > \lambda
_{F} $ and $n_{max}
> > 1$, the size quantization is particularly insignificant and
this case corresponds to the classical mechanics approach. From
Eqs.(\ref{eq19})-(\ref{eq23}) we obtain

\begin{equation}
\label{eq24} \chi = \left\{ {{\begin{array}{*{20}c}
 {{\begin{array}{*{20}c}
 {\frac{{2mer\left( {0} \right)V}}{{\pi ^{2}\hbar ^{2}\Phi} },} \hfill &
{|\theta | < \Phi /2,} \hfill \\
\end{array}} } \hfill \\
 {} \hfill \\
 {{\begin{array}{*{20}c}
 {0,} \hfill & {|\theta | > \Phi /2.} \hfill \\
\end{array}} } \hfill \\
\end{array}} } \right.
\end{equation}

\begin{figure}
\includegraphics[height=6 cm,width=6 cm]{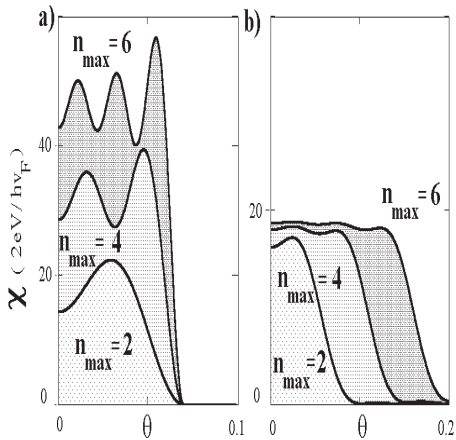}
\caption{The angular ($\theta$ in radians) dependence of the
electron flow (sum over all conducting modes) from a constriction,
corresponding to $n_{max} = 2,4,6$. Results are shown for: (a) a
constriction with a shape close to a linear widening one, and (b)
a constriction that ends abruptly. The parameters of the
constrictions are as in Fig.{\ref{fig3}}. } \label{fig5}
\end{figure}

\begin{figure}
\includegraphics[height=6 cm,width=6 cm]{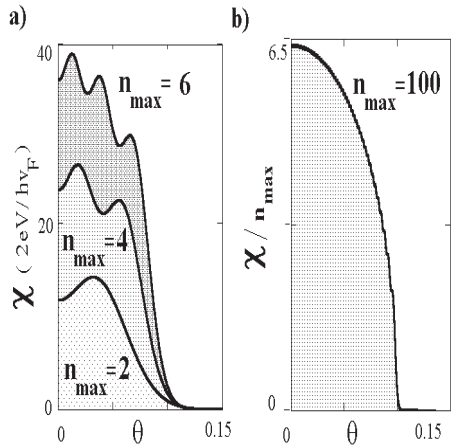}
\caption{The angular ($\theta$ in radians) dependence of the
electron flow (sum over all conducting modes) from a constriction.
Results are shown for: (a) $n_{max} = 2,4,6$ and (b) $n_{max} =
100$ (normalized), for a soft-wall model. The parameters of the
constriction are as in Fig.{\ref{fig3}}.} \label{fig6}
\end{figure}

We observe that if $n_{max} $ is not too large, the electron beam
distribution $\chi$ oscillates with a period $\Phi /n_{max} $ and
the amplitude of the oscillation grows at the edges of the flow at
$\theta = \pm \Phi /2$ (see Fig.{\ref{fig5}a}, $n_{max} = 2,4,6$).
The summation of the contributions of different modes radiated by
the constriction which ends abruptly (Eqs.(\ref{eq21}),
(\ref{eq22})) gives a result similar to Eq.(\ref{eq24}) with
additional numerically small oscillations (see Fig.\ref{fig5}b). A
$\theta$-dependence of the beam distribution that is similar to
Eq.(\ref{eq24}) has been predicted in Ref.\onlinecite{b3}. Note
that the "step-like" dependence, with sharp edges at $\theta =
\Phi /2$, is not universal. It takes place only in the classical
limit for both types of constrictions discussed above. In
Fig.{\ref{fig6}} we present also the $\theta$-dependence of the
beam distribution for the soft-wall model corresponding to
different values of $n_{max} $. Apparently, in the classical
limit, the angular distribution of the radiated electron beam that
is generated by a constriction with a shape described by the
expression $r \propto z^{\alpha} $ for $\alpha > 1$ (at least up
to the detachment point), has no sharp edges at $\theta = \pm \Phi
/2$.

\section{CONCLUSION}

The analysis that we performed demonstrates that extremely narrow
electron beams may be generated by a voltage applied to
sufficiently long narrow constrictions. The minimal length $L$ of
such a constriction is related to the minimal half-width, $r(0)$,
and the angular size of the beam, $\Phi$, through Eq.(\ref{eq9}).

An alternative scheme for generation of a super-narrow electron beam may be
achieved by a specially tuned electrostatic potential applied to a
sufficiently wide constriction, in juxtaposition with blocking of all the
electronic size quantization modes in the constriction, except for the
lowest one (here, the minimal width of constriction has to be much larger
than the electron wave length). To minimize the ``integral'' width of the
beam, which combines its angular and spatial widths, one should use
constrictions made of conducting materials with high electron densities.

We have also illustrated here that the angular distribution of the
electron density in the beam provides information about the
quantum adiabatic electronic states inside the constriction. When
the adiabatic region ends smoothly, the electron density in the
beam reproduces the probability density in the coordinate
representation. This result elucidates the feasibility condition
for the electron flow distributions observed in Ref.
\onlinecite{a3}-\onlinecite{a6} and \onlinecite{a8} - accordingly,
the radius of curvature of the constriction in the detachment
point should be larger than $r_{t} /\Phi ^{2}$. If the adiabatic
region ends abruptly the electron density in the beam reproduces
the probability density in the momentum representation.

\section{Acknowledgements}

This research was made possible in part by Grant No.UP2-2430-KH-02 of the
U.S. Civilian Research \& Development Foundation for the Independent States
of the Former Soviet Union (CRDF). The research of E.N.B and U.L. was also
supported by the US Department of Energy, Grant No. FG05-86ER-45234.

\end{document}